# Performance of Titanium-Oxide/Polymer Insulation in Bi-2212/Ag-alloy Round Wire Wound Superconducting Coils

**Peng Chen[1, 2], Ulf P Trociewitz[1], Matthieu Dalban-Canassy[1], Jianyi Jiang[1], Eric E Hellstrom[1, 2] and David C Larbalestier[1, 2]**

[1] Applied Superconductivity Center, National High Magnetic Field Laboratory, Florida State University, 2031 East Paul Dirac Drive, Tallahassee, FL 32310, USA

[2] Department of Mechanical Engineering, FAMU-FSU College of Engineering, Florida State University, 2525 Pottsdamer Street, Tallahassee, FL 32310, USA

Email: pengchen@asc.magnet.fsu.edu

**Abstract**

Conductor insulation is one of the key components needed to make Ag-alloy clad $Bi_2Sr_2CaCu_2O_{8+x}$ (Bi-2212/Ag) superconducting round wire (RW) successful for high field magnet applications as dielectric standoff and high winding current densities ($J_w$) directly depend on it. In this study, a $TiO_2$/polymer insulation coating developed by nGimat LLC was applied to test samples and a high field test coil. The insulation was investigated by differential thermal analysis (DTA), thermo-gravimetric analysis (TGA), scanning electron microscopy (SEM), dielectric properties measurement, and transport critical current ($I_c$) properties measurement. About 29% of the insulation by weight is polymer. When the Bi-2212/Ag wire is full heat treated, this decomposes with slow heating to 400°C in pure $O_2$. After the full reaction, we found that the $TiO_2$ did not degrade the critical current properties, adhered well to the conductor, and provided a breakdown voltage of >100 V, which allowed the test coil to survive quenching in 31.2 T background field, while providing a 2.6 T field increment. For Bi-2212/Ag RW with a typical diameter of 1.0-1.5 mm, this ~15 μm thick insulation allows a very high coil packing factor of ~0.74, whereas earlier alumino-silicate braid insulation only allows packing factors of 0.38-0.48.

## 1. Introduction

Ag-alloy clad $Bi_2Sr_2CaCu_2O_{8+x}$ (Bi-2212/Ag) multi-filamentary round wire (RW) is fabricated by the powder-in-tube (PIT) method, yielding long piece lengths suitable for magnet construction. Bi-2212/Ag RW exhibits a high irreversibility field beyond 100 T and a high critical current density exceeding $10^5$ A cm$^{-2}$ at 4.2 K up to 45 T [1, 2]. As the only high temperature superconductor (HTS) with RW geometry, Bi-2212/Ag RW has the advantage of being multi-filamentary and without macroscopic anisotropy, making it rather attractive for high field magnet construction. Being a round wire, it can be cabled easily, which makes it particularly appropriate for accelerator applications where Rutherford cables are preferred [3-6]. Recently, a Bi-2212/Ag RW wound superconducting coil successfully achieved 33.8 T combined field when tested in a 31.2 T resistive magnet at the National High Magnetic Field Laboratory (NHMFL) [7]. In contrast to highly aspected tape conductors, Bi-2212/Ag RW is not expected to be significantly affected by shielding currents that are induced by the large radial field components that generally occur at coil ends [8-11]. This isotropic behavior may be an important benefit for its use in high-field, high-homogeneity magnetic field applications, such as for nuclear magnetic resonance (NMR) magnets [12, 13].

For magnet applications, effective conductor insulation is crucial to provide sufficient electrical standoff within the coil winding pack. Besides good dielectric properties, the coating should be thin, since insulation space dilutes the overall winding current density ($J_w$), particularly in single strand coils. Since reacted Bi-2212 is a brittle ceramic, coils are typically built applying the Wind and React (W&R) approach [14, 15]. The insulation thus has to be applied to the conductor before coil winding. Subsequent to winding, Bi-2212/Ag coils are heat treated in flowing pure oxygen at a maximum temperature of around 890 °C to achieve their superconducting properties. These conditions pose several challenges for an effective insulation for Bi-2212/Ag, which must be chemically compatible with Bi-2212/Ag round wires, in particular with the Ag-alloy with which it is in contact and whose electrical conductivity must not be degraded since it serves as the electrical stabilizer for quench protection. The insulation must be permeable to oxygen, to allow both for oxygen loss during the partial melting of the Bi-2212 and subsequent oxygen uptake during solidification and full development of the overdoped state during cooling. Lastly the insulation has to be abrasion resistant to allow for handling during coil winding and thin enough to permit a high winding current density $J_w$ without sacrificing dielectric strength.

Currently, commercial Bi-2212/Ag RW is insulated using a rather thick ceramic fiber wrap or braid. Wire manufactured by Oxford Superconducting Technology (OST) is insulated with a braided 73/27 wt% alumino-silicate fiber sleeve made of crystalline alumina and amorphous silicate. This insulation, however, is not adequate for Bi-2212/Ag RW for two main reasons. First, previous research indicates that alumino-silicate braid insulation reacts with Bi-2212/Ag RW [16], allowing silver to be absorbed into the braid fibers, as shown in figure 1. This reaction erodes and thins the Ag-alloy sheath, potentially exacerbating Bi-2212 leakage during the partial melt heat treatment (HT) [16-18]. Comparing short bare samples with alumino-silicate braid-insulated samples processed under nominally identical conditions, the critical current ($I_c$) of alumino-silicate braided samples is typically 15-20% lower than that of bare samples [19]. Previous studies found that the oxygen content of the conductor (*i.e.* the doping state) is not affected by the alumino-silicate braid insulation [19, 20], avoiding this as a possible variable. A second drawback of the alumino-silicate braid insulation is that it is about 150 μm thick, thus imposing a significant loss of conductor packing factor in the coil, whereas the $TiO_2$-based insulation studied here is around 15 μm, one order of magnitude thinner. With these potential advantages in mind, we decided to make a full evaluation of the insulation, both in small multi-layer spirals and then in a coil that safely generated more than 2 T when tested in the 31.2 T magnet at the NHMFL [7]. We find in fact that there are many advantages to this new insulation that confer real benefit to high field magnet use.

## 2. Experimental details

Two lengths of Bi-2212/Ag multi-filamentary RW were insulated by nGimat LLC with $TiO_2$/polymer before HT. Both wires were fabricated by OST using the PIT method with an outer sheath of Ag-0.2 wt% Mg alloy. For both short and spiral samples we used an insulated wire of ~0.83 mm diameter and for small test coils a ~1.43 mm diameter insulated wire. Initially, short and spiral samples were heat treated using the standard HT schedule under 1 bar of flowing oxygen, as detailed in figure 2. After studying the thermal behavior of the $TiO_2$/polymer coating by differential thermal analysis (DTA) and thermo-gravimetric analysis (TGA), we realized that heating rates in the range up to ~400°C had to be slowed in order to prevent coating flaking off the wire. We tested and analyzed several spirals using slower ramps to determine the optimum HT for coil processing.

The surface morphology and thickness of the insulation were examined by scanning electron microscopy (SEM) and the dielectric strength was evaluated by DC breakdown voltage at room temperature (RT) using a megohmmeter fed from two cylindrical electrodes placed in close contact with various sections of insulated Bi-2212/Ag RW. A qualitative evaluation of the mechanical strength and pliability of the insulation coating in both unreacted and reacted states was microscopically compared with twisted wire pairs with various twist pitches before the HT.

To understand the surface morphology evolution of the insulation during the HT, several identical single-layer spiral samples cooled from various stages in the HT were generated. These spirals were wound on 1.5 cm diameter Inconel 600 alloy mandrels. To reduce chemical interaction between the samples and the Inconel, the mandrels were pre-oxidized and spray coated with zirconia prior to winding. Each spiral had about 0.5 m of wire wound in 10 turns. Based on the DTA and TGA analysis of the insulation, several key temperature points were chosen for study. The heating rate to reach these points was 160°C/h starting from RT, after which the furnace was cooled to RT all the while flowing oxygen at 1 l/min. After cool down, sample surface morphology was studied by SEM, and breakdown voltages were measured on several different sections of each sample.

To mimic coil behavior, several tightly wound multi-layer spirals (using about 3 m of wire in 6 layers, each with about 10 turns) were made on the same Inconel 600 mandrels. Some spirals used slower ramps in different temperature ranges below 400°C suggested by the DTA and TGA results, while others were processed using the standard route with faster ramps as illustrated in figure 2. Their surface morphology was then compared.

To understand to what extent the coating affects the transport properties of the wire, insulated and bare short samples were heat treated identically, and $I_c$ measurements were performed using a standard four-point method with a 1 µV/cm criterion at 4.2 K and 5 T background magnetic field. To investigate variations of the transport properties across the different layers in the multi-layer spirals, the middle turns of every spiral layer were extracted.

### 3. Results

#### 3.1. Fundamental properties of the TiO$_2$/polymer insulation

Figure 3 shows an SEM image of the transverse cross section of a short sample in the green state (*i.e.* before the HT). The insulation is indeed much thinner than the alumino-silicate braid used earlier, with a circumferentially uniform thickness of ~15 µm. The insulation has a double-layer structure. The ~13 µm thick base layer contains TiO$_2$ nanoparticles in a polymer. To improve handling of the wire, an additional ~2 µm layer made of the same polymer is applied on top of the first layer. During the standard HT of figure 2, the polymer burns off between 200-300 °C, leaving the TiO$_2$ sintered to the wire as shown in figure 4. Figure 5 shows the insulation surface morphology after the standard HT of figure 2 and the optimum HT with a 20°C/h slow ramp rate from RT to 400°C, respectively. In both situations, it is clear that a rather dense network of cracks that reach through the oxide to the Ag-alloy wire sheath is quite visible. These cracks are typically 10-15 µm wide and occur more often on the tensile side of the wire bend after winding on a coil form. While cracks are in general not beneficial to the dielectric properties of the coating, they are almost certainly beneficial to the Bi-2212 phase quality, since optimum superconducting properties require oxygen transport through the coating and Ag-alloy matrix during the HT.

Figure 6 shows light microscope images of twisted wire pair samples using ~0.83 mm diameter insulated wire with different pitch lengths before and after the full, standard HT illustrated in figure 2. For pitch lengths not greater than 5 mm, the coating buckles and cracks even before the HT, as shown in figure 6(a). For twist pitches equal 7 mm, however, while the unreacted twisted sample looks pristine without any buckling and cracking, cracks appear in the coating after the full HT, as shown in figure 6(b). If the pitch length of the unreacted sample is increased further to 15 mm, the coating surface stays smooth and does not show signs of flaking after the full HT, which indicates good adhesion of the coating, as shown in figure 6(c).

#### 3.2. Thermal analysis of the TiO$_2$/polymer insulation

To understand the thermal behavior of the insulation, simultaneous DTA and TGA were carried out on pieces mechanically extracted from coated wires with 5 °C min$^{-1}$ heating ramp rate in flowing oxygen. The DTA data, shown in figure 7, reveal an endothermic peak at ~180 °C corresponding to the melting of the insulation, while the exothermic peak at ~260 °C corresponds to the burn-off of the polymer in the coating. The TGA curve of figure 7 illustrates a continuous mass loss of the coating mainly between RT and 400 °C, being particularly large between 100 °C and 230 °C.

### 3.3. Furnace cooling experiments

On the basis of the DTA and TGA data of figure 7, six key temperatures were identified for a series of intermediate process furnace cooling experiments. Figure 8 shows SEM images of the insulated wire surfaces of these samples. At 100 °C, there are no cracks in the coating, and the coating is gooey, potentially resulting from the beginning of decomposition of the polymer. At 150 °C, the wire surface is still smooth but the coating surface has become shiny. Between 200 °C and 520 °C, the surface gradually turns gray-green and cracks of greater density start to develop, but overall the surface coverage remains complete and rather uniform. We found that the breakdown voltage and the insulation weight loss curve are very well correlated, as shown in figure 9. A marked decrease of the breakdown voltage from ~1200 V to 150 V occurs in the temperature range of ~100-230°C, where ~27% of the insulation mass is lost. The insulation breakdown voltage does not degrade further after the full HT, where the entire polymer is burned out and the final insulation mass loss equals ~29%.

### 3.4. Multi-layer spirals

Several Bi-2212/Ag RW wound multi-layer spirals with ~3 m of insulated wire consisting of two separately terminated layer sets wound tightly on top of each other were manufactured so as to allow the dielectric properties between the two layer sets to be measured. Two multi-layer spirals with the same structure were heat treated with slow ramps in different low temperature regimes to protect the insulation. One multi-layer spiral (Spiral A) was heat treated using the standard route of figure 2 except for applying a 20°C/h slow ramp rate between 200°C and 260°C, which covered only a portion of the temperature range during which the polymer burns off. Figure 10(a) shows the typical surface morphology of the innermost layer of Spiral A, which clearly shows some insulation flaking. Another multi-layer spiral (Spiral B) was heat treated using the standard route of figure 2 except for applying a 20°C/h slow ramp over an extended range from RT to 400°C, where the TGA curve of figure 7 shows most of the weight loss to occur. Figure 10(b) shows the surface morphology of a sample extracted from the innermost layer of Spiral B, where there were some insulation cracks but no flaking. Comparing figures 10(a) and 10(b), it becomes clear that a slow ramp rate is crucial from RT to 400 °C during the HT.

Figure 11 shows the $I_c$ values (4.2 K, 5 T) of samples extracted from each layer of Spiral B. Spiral B shows homogeneous transport properties throughout all the layers, with an average $I_c$ of 167 A and a standard deviation of 3 A. For comparison, transport data of two short samples (bare and insulated) are included in figure 11 as well. The short insulated sample has an $I_c$ of 207 A, while that of the short bare sample has almost the same $I_c$ of 209 A. These almost identical transport properties for both short samples indicate that the $TiO_2$/polymer insulation is chemically compatible with Bi-2212/Ag RW and does not degrade the wire composition during the HT. We attribute the difference in short and long-length $I_c$ to the usual length-dependent degradations explored recently in detail by Malagoli *et al* [17, 18] and conclude that this effect occurs independent of the presence or absence of the $TiO_2$-based insulation.

### 3.5. Insulation of a small solenoid coil

Figure 12 shows a coil insulated with the $TiO_2$/polymer coating after each of three different key processing stages: after coil winding, after insulation burn-off and after 10 bar overpressure (OP) processing, which is described below. The image clearly suggests that the insulation is unaffected by coil winding, as shown in figure 12(a), which in this case equals about 9-25% bending strain. Due to improper handling during winding, however, some insulation coating did flake off the outmost layer of the coil. As indicated by the multi-layer spiral study, it is essential to apply slow heating to remove the polymer in the insulation. Accordingly the coil was ramped up from RT to 400 °C at ~20°C/h before it was cooled to RT for transferring it to OP furnace. After insulation burn-off, the initially brown insulation turned gray-green, as presented in the furnace cooling experiments. However, no additional insulation coating flaking was observed, as shown in figure 12(b). For the 10 bar OP processing procedure the two ends of the coil were sealed with Ag and a reaction gas mixture of 1 bar oxygen and 9 bar argon was applied during the typical HT schedule of figure 2. Because the polymer had already been burned out, the 160 °C/h heating rate shown in figure 2 was used from RT to 821 °C.

After the complete HT the insulation turned brown in color, but still did not show any signs of flaking during the coil OP processing, as illustrated in figure 12(c). The wire diameter of the coil remained ~1.4 mm and the wire showed no leakage, which means that the 9 bar Ar pressure prevented expansion of the wire during the HT. The coil resistance remained constant at ~0.25    through all three processing steps, which indicates that there were no electrical shorts in the coil.

This coil generated a combined magnetic field of 33.8 T in 31.2 T background at 1.8 K cryogenic bath temperature at the NHMFL and was quenched multiple times without damage [7]. The Hall sensor showed a linear relationship with the coil operating current at all times, which also indicates that the insulation provided sufficient dielectric standoff.

## 4. Discussion

In this study, the performance of a $TiO_2$/polymer insulation coating applied by nGimat LLC was evaluated. The insulation appears to be robust through the HT, it was found to be chemically compatible with Bi-2212/Ag RW, and it did not degrade the wire performance. Since this insulation is only ~13 µm thick after the HT, it is highly beneficial in increasing the winding current density $J_w$ in coils since it significantly increases the coil packing density. A graphical comparison of packing factors between the 15 µm thick $TiO_2$/polymer insulation and the much thicker (150 µm) alumino-silicate braid insulation for Bi-2212/Ag RW with different diameters is illustrated in figure 13. It is clear that the coil packing factor can be generally increased from 0.38-0.48 using the alumino-silicate braid insulation to ~0.74 using thin $TiO_2$-based insulation, especially for 1.0 mm diameter Bi-2212/Ag RW, the coil packing factor can be increased dramatically from 0.38 to ~0.74.

The DTA and TGA data shown in figure 7 provide valuable information for HT optimization of $TiO_2$/polymer insulated Bi-2212/Ag superconductors. The data have shown that it is essential to apply a slow ramp rate to smoothly remove the polymer in the insulation. The data also clearly define temperatures corresponding to major changes in the insulation which we therefore studied by furnace cooling samples at these points. As shown in the TGA curve of figure 7, the insulation continuously loses mass between RT and ~400 °C during decomposition of the polymer, finally losing ~29% of the mass. The mass loss leads to gaps between turns and layers, which may lead to coil sagging during the HT. However, due to the fact that the top layer of the insulation is just ~2 µm thick, this risk may be considered minor and can be further mitigated if a stiffer or larger diameter Bi-2212/Ag conductor is selected for coil winding.

A small Bi-2212/Ag solenoid coil coated with the $TiO_2$/polymer insulation was built at the NHMFL. As shown in figure 12(a), there was some insulation damage at the outmost layer of the coil due to improper handling, but we conclude that the insulation was not affected by the 9-25% bending strain applied in this coil. Following the multi-layer spiral tests shown in figure 10, a slow ramp of ~20°C/h was applied from RT to 400 °C for the polymer decomposition. No insulation coating flaking happened during this processing step, as is shown in figure 12(b). Kametani et al [21] have recently reported that void space and bubble formation are the chief reasons for wire expansion and leakage, which severely reduce the transport properties of Bi-2212/Ag RW, while the 10 bar OP processing during the HT of this coil effectively prevented dedensification of the wire and improved the transport properties. Jiang et al [22], who recently developed the highly effective OP heat treatment processing for Bi-2212 conductor, have shown that Bi-2212 conductor can be further densified by increasing the pressure of the OP processing, which will further improve transport properties. As shown in figure 12(c), the coil did not show any signs of insulation coating flaking resulting from OP processing, which clearly indicates this insulation is compatible with the OP processing. Note that the slight damage to the coating visible in figure 12 occurred during coil winding, but it did not compromise the coil which generated a combined magnetic field of 33.8 T in 31.2 T background at the NHMFL. We believe that this is a valuable demonstration of the successful application of this $TiO_2$/polymer insulation.

In addition to the commercial $TiO_2$/polymer insulation discussed in this paper, we have investigated two other different routes for Bi-2212/Ag RW insulation. We investigated sol-gel based $ZrO_2$ and $Al_2O_3$ coatings through collaboration with Harran University, Turkey [23-26]. Sol-gel technology is widely used and advantageous for its ease of application and potential low cost [27]. However, a persistent lack of adhesion on Ag and the need for multiple dips to build up a thick enough insulation for Bi-2212/Ag RW makes us, presently at least, doubt this sol-gel approach is viable for long length coating. We are presently evaluating another ceramic-polymer based insulation coating at the NHMFL [28].

## 5. Summary

In this paper, we present an investigation of the $TiO_2$/polymer insulation being offered by nGimat LLC as a potential insulation solution for Bi-2212/Ag RW wound superconducting coils. The insulation has a double-layer structure, the base layer consisting of $TiO_2$ nanoparticles in a polymer with a thickness of ~13 μm. To improve handling of the wire, nGimat LLC applies a top layer that is ~2 μm thick made of the same polymer. During the HT, the polymer decomposes and burns off, leaving the $TiO_2$ sintered onto the wire. After the HT, the insulation surface exhibits a rather dense network of cracks with typical width of 10-15 μm. These cracks appear to allow oxygen permeability to the underlying wire, which is necessary for Bi-2212 phase formation. The breakdown voltage of this insulation is ~1200 V before the HT and ~150 V after the full HT, explained by the appearance of the network of cracks after the HT. DTA shows one endothermic peak at ~180 °C and one exothermic peak at ~260 °C. The endothermic peak corresponds to the melting of the insulation, while the exothermic peak corresponds to the burn-off of the polymer in the coating. The TGA data clearly show continuous mass loss between RT and ~400 °C, the majority of which occurs between 100 °C and 230 °C. The mass loss corresponds to the decomposition of the polymer, the final mass loss percentage being ~29%. The TGA data also suggest that it is essential to apply a slow ramp rate to remove the polymer. The transport properties of insulated and bare samples were essentially identical, which indicates chemical compatibility with Bi-2212/Ag RW. A Bi-2212/Ag RW wound solenoid coil was built using this insulation. The coil resistance was constant through coil winding, polymer burn-off and full coil reaction. The coil was successfully tested at the NHMFL generating 33.8 T combined magnetic field in a 31.2 T background field [7]. Multiple quenches occurred safely, which also illustrates that the insulation provided sufficient dielectric standoff.

## Acknowledgements

We are grateful for discussions with members of the Very High Field Superconducting Magnet Collaboration (VHFSMC). We express our thanks to Youri Viouchkov from the NHMFL Magnet Science and Technology department for CAD work on the spiral mandrels and coordinating coil part manufacturing. This work was supported by an ARRA grant of the US Department of Energy Office of High Energy Physics and by the National High Magnetic Field Laboratory which is supported by the National Science Foundation under NSF/DMR-1157490 and by the State of Florida.

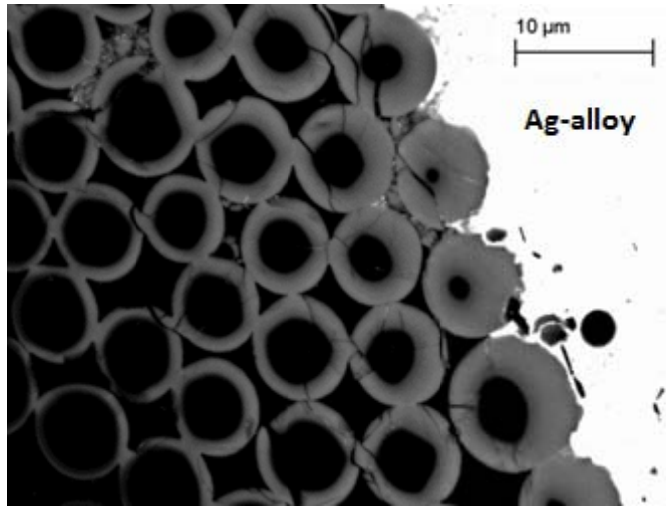

Figure 1. SEM image of a transverse cross section of the alumino-silicate braid insulation around a fully processed Bi-2212/Ag RW. The gray circles on the fibers result from the chemical reaction of Ag with the alumino-silicate braid. The fiber cracks indicate the poor mechanical strength of the braid insulation after the HT.

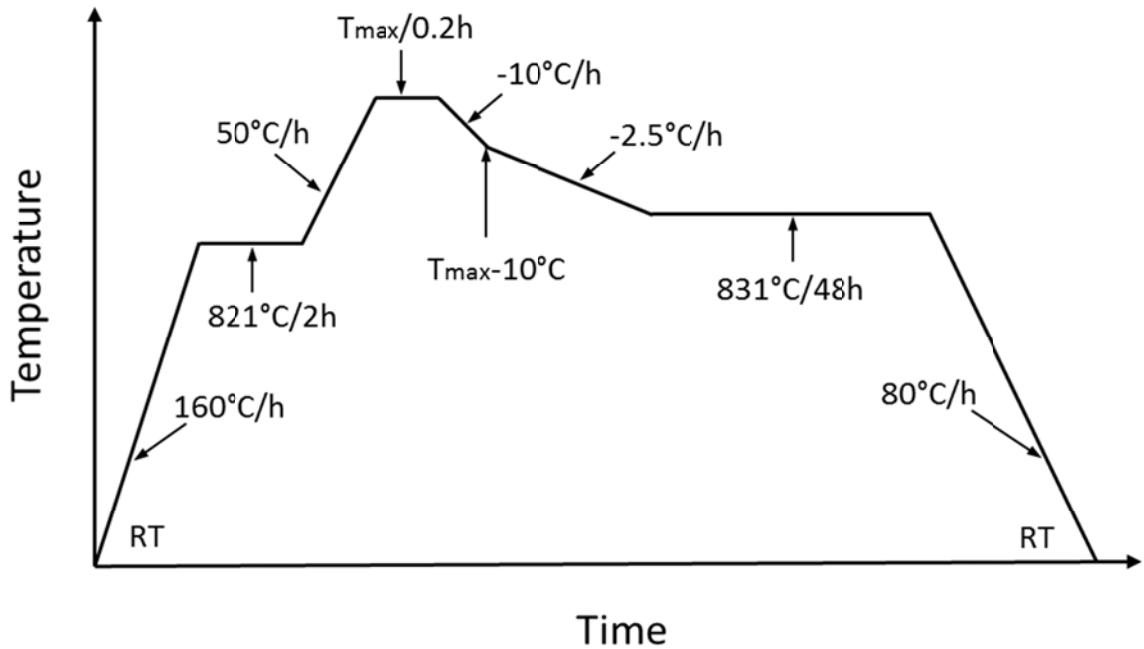

Figure 2. Typical standard HT profile used for Bi-2212/Ag RW, including the insulation experiments performed here. Note that optimum processing of the $TiO_2$/polymer insulated conductor replaces the initial ramp from RT to 400°C with a slow ramp of 20°C/h that allows smooth decomposition of the polymer.

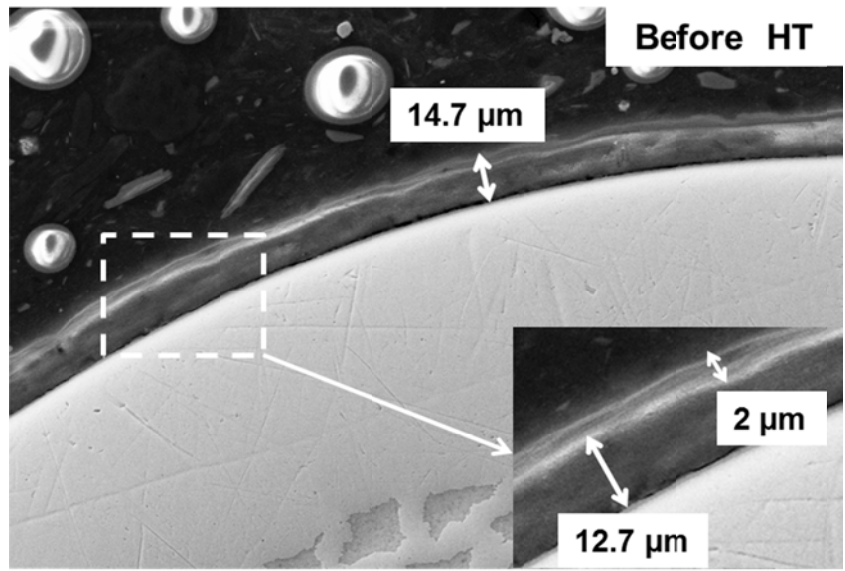

Figure 3. SEM image of Bi-2212/Ag RW with the $TiO_2$/polymer insulation before the HT showing that the insulation consists of two layers. The base layer, with a typical thickness of about 13 μm, is made of $TiO_2$ nanoparticles and a polymer, while the top layer, used to improve winding durability, is composed of the same polymer with a typical thickness of about 2 μm.

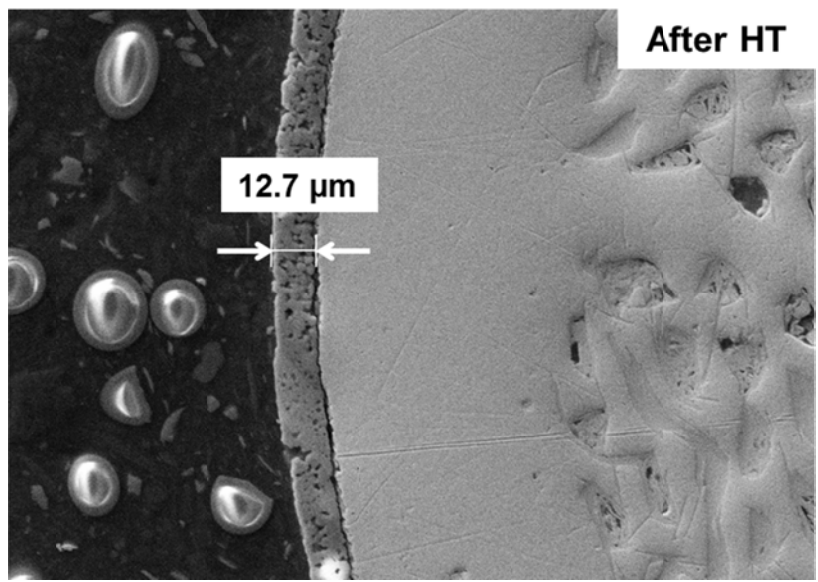

Figure 4. SEM image of Bi-2212/Ag RW with the $TiO_2$/polymer insulation after the standard HT of figure 2, showing that the volatile top layer of polymer has burnt off, leaving an adherent base layer of sintered $TiO_2$ particles with a uniform thickness of about 13 μm. The holes in the $TiO_2$ layer suggest the polymer has burned out of the base layer too. DTA/TGA data do not show that the polymer burns out of the top layer and base layer at different temperatures.

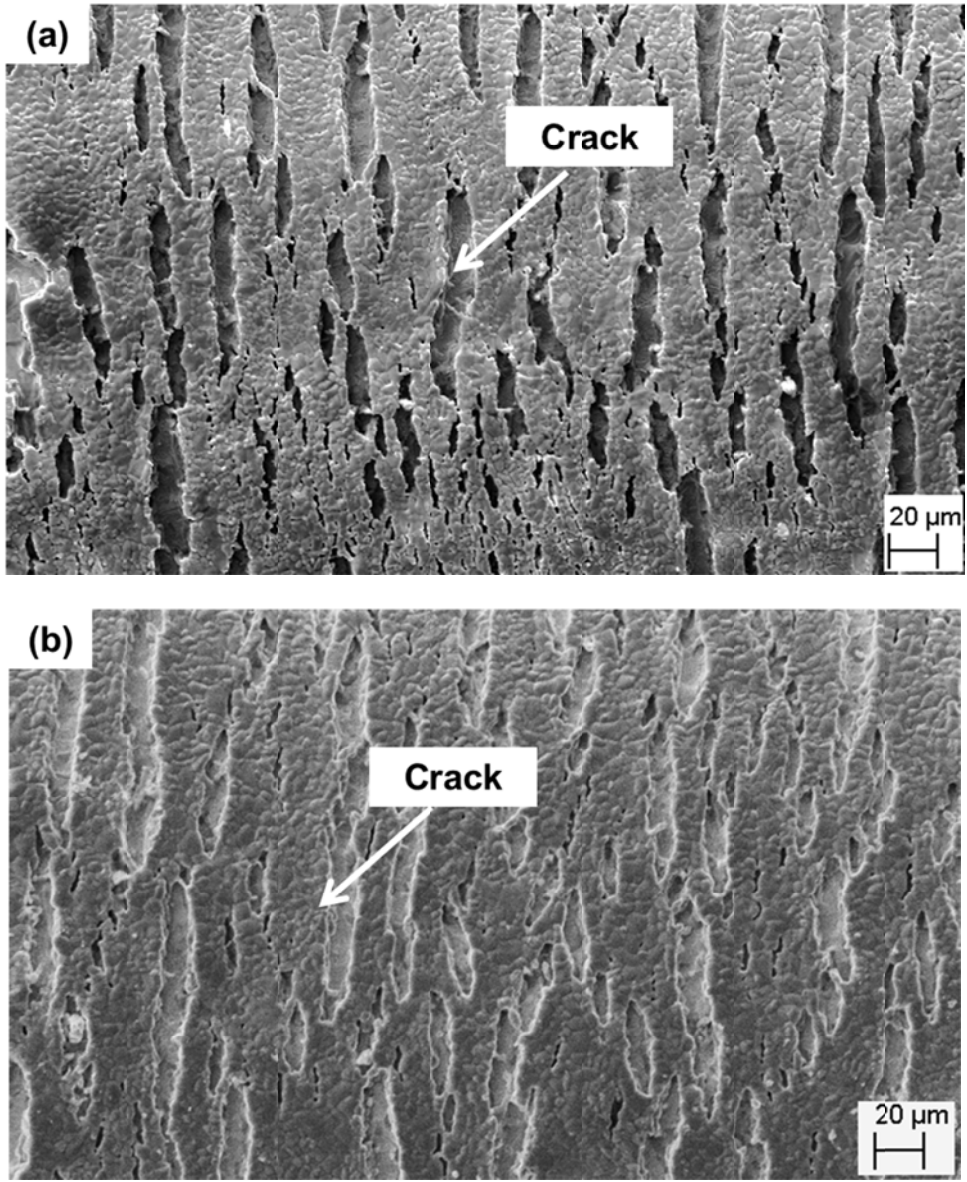

Figure 5. SEM surface images of the insulation, seen (a) after the standard HT of figure 2 and (b) after the optimum HT with a 20°C/h slow ramp rate from RT to 400°C. Both images present cracks through the insulation, which may provide a crucial oxygen pathway to the Bi-2212/Ag RW during HT.

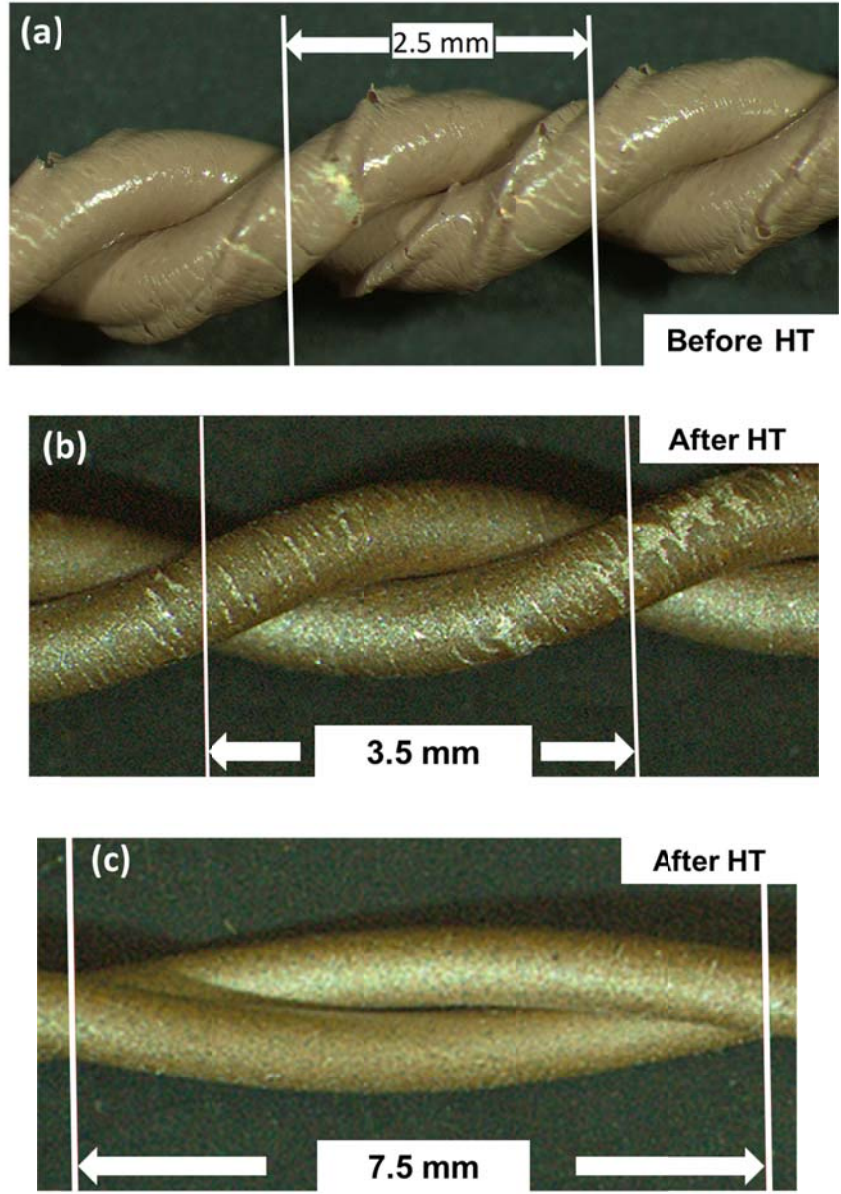

Figure 6. Light microscope images of twisted pair samples of insulated 0.83 mm diameter wire showing (a) some buckling and cracks in the coating of unreacted samples wound with a half pitch length of 2.5 mm, (b) for samples wound with a half pitch length of 3.5 mm, after the standard HT of figure 2, there are some cracks and coating flaking, and (c) for samples wound with a half pitch length of 7.5 mm, after the standard HT of figure 2, the sample surface is smooth without any coating flaking.

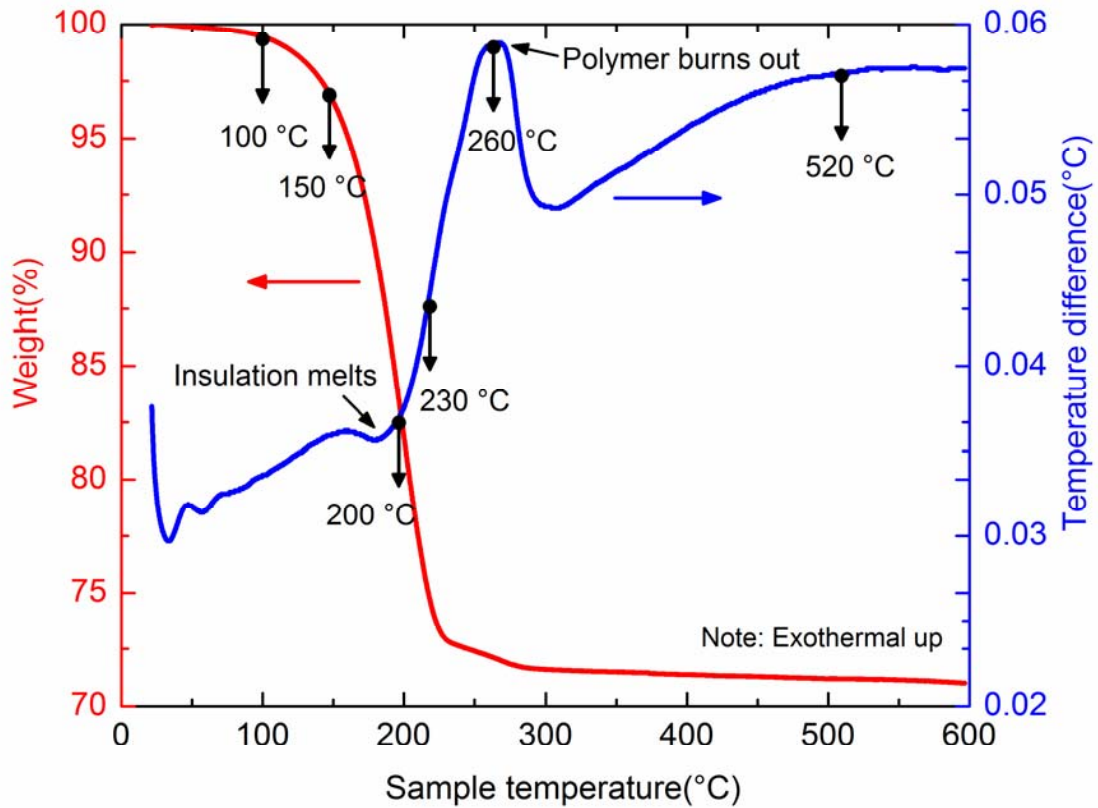

Figure 7. DTA and TGA curves of the TiO$_2$/polymer insulation which illustrate partial insulation melting at about 180 °C, and the polymer burns out starting at about 260 °C. The TGA curve illustrates that the insulation continuously loses mass mainly between RT and 400 °C, where about 29% of the insulation mass is lost. Temperatures (100°C, 150°C, 200°C, 230°C, 260°C and 520°C) from which samples were furnace cooled and examined in figure 8 are indicated by the arrows.

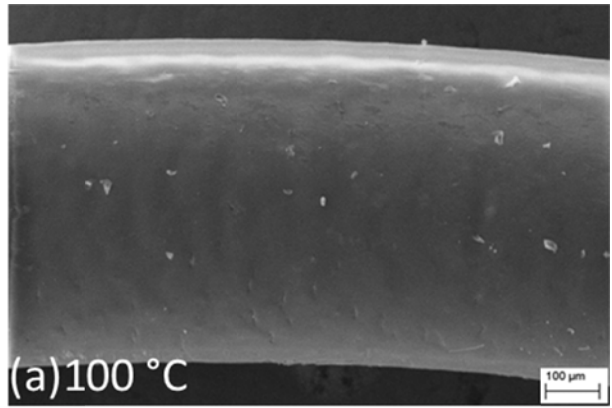

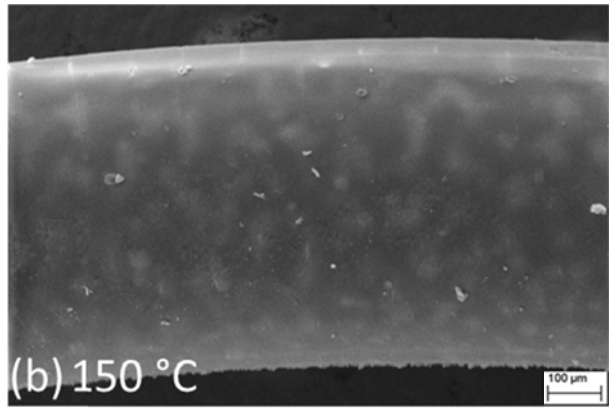

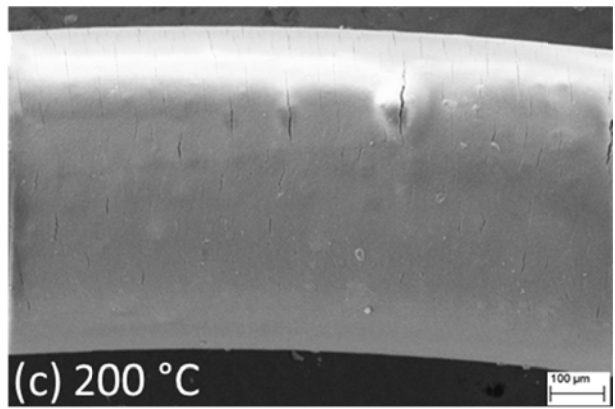

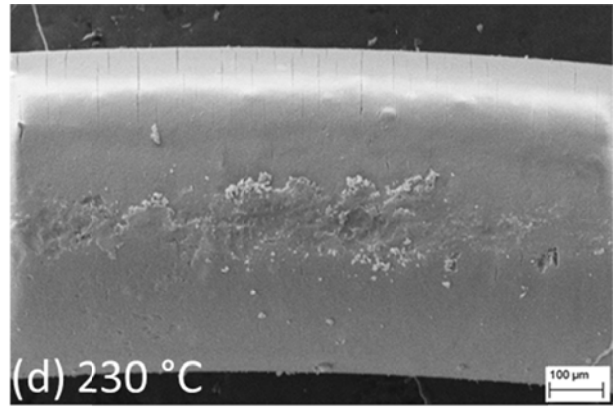

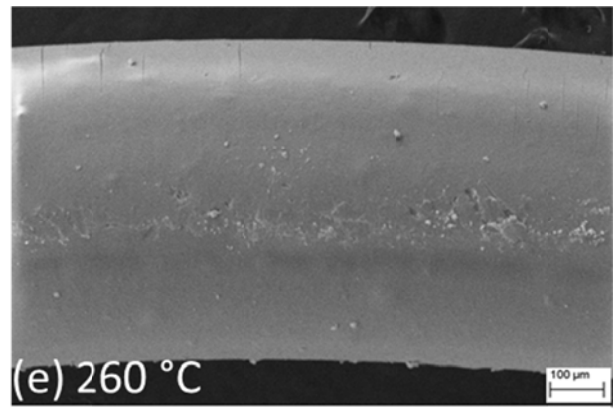

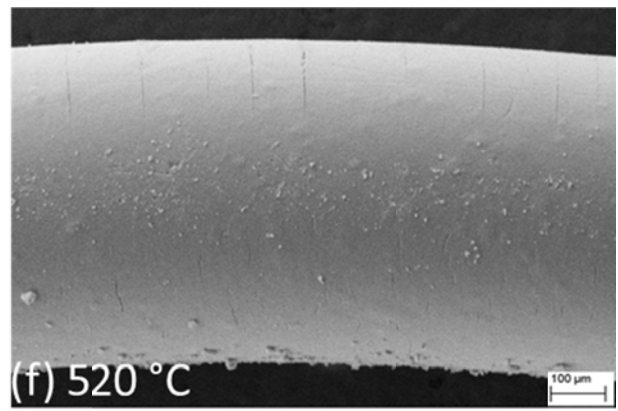

Figure 8. SEM surface images of samples furnace cooled from (a) 100 °C (b) 150 °C (c) 200 °C (d) 230 °C (e) 260 °C and (f) 520 °C. Insulation cracks developed between 200 °C and 520 °C, but the insulation was well preserved in general.

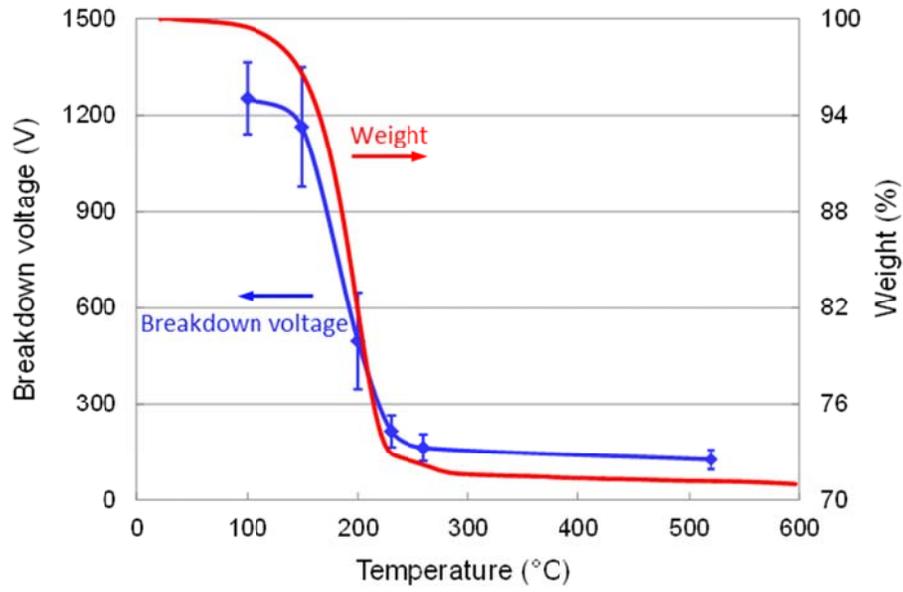

Figure 9. Breakdown voltage, measured over a range of at least 6 samples, after furnace cooling from different temperatures. With increasing furnace cooling temperature, the sample breakdown voltage decreases, scaling well with the loss of polymer seen in the TGA curve of figure 7.

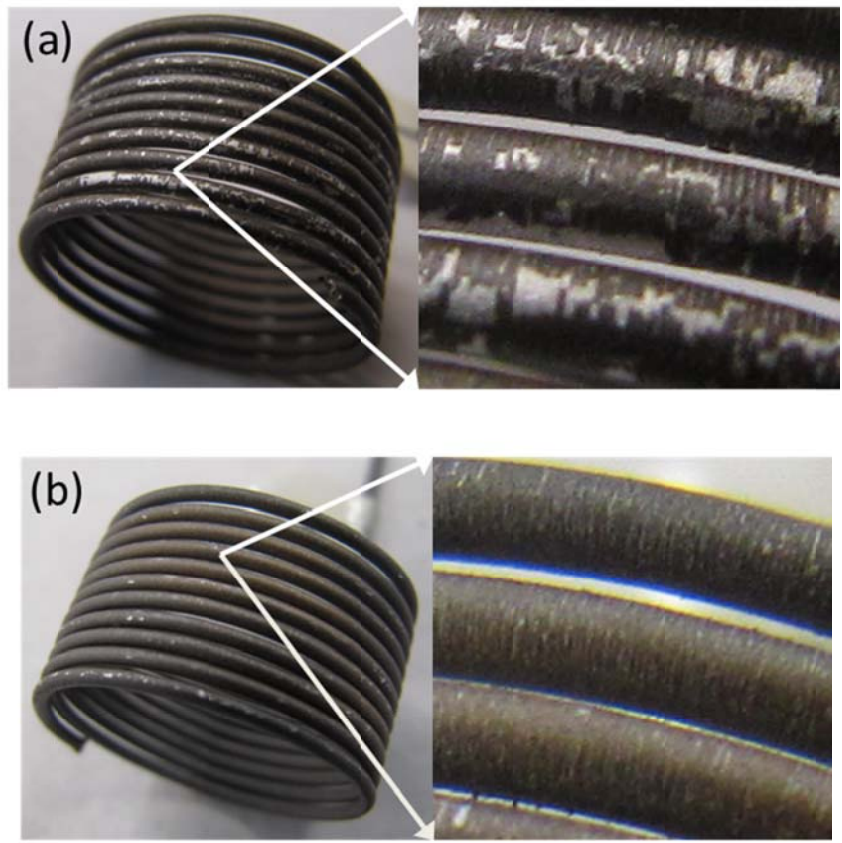

Figure 10. (a) Image of the innermost layer of one multi-layer spiral (Spiral A), heat treated with the standard HT schedule shown in figure 2, except for a 20°C/h slow ramp rate from 200°C to 260°C over the major polymer burning off region observed in the DTA curve of figure 7. After the full HT, there is some evidence of local insulation flaking. (b) Image of the innermost layer of another multi-layer spiral (Spiral B), heat treated using the standard HT schedule shown in figure 2, except for a 20°C/h slow ramp rate from RT to 400°C over the major weight loss region observed in the TGA curve of figure 7. After the full HT, there are some surface insulation cracks but now there is no coating flaking.

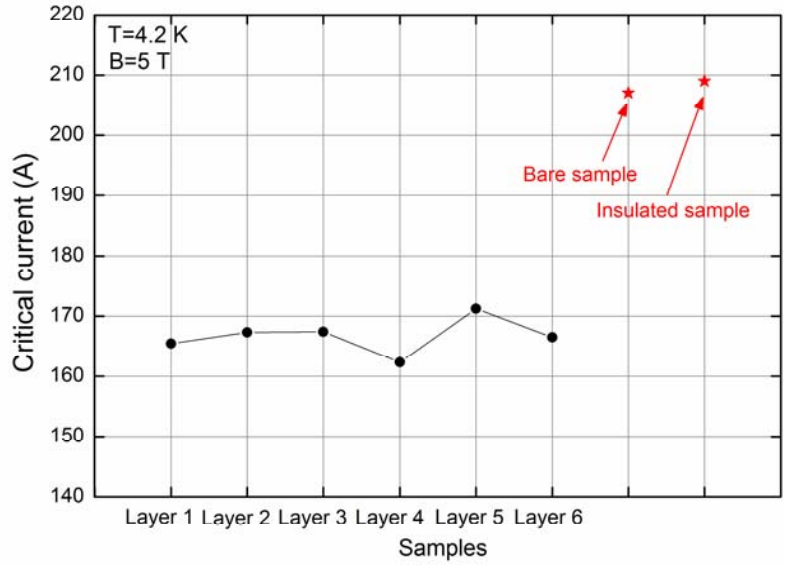

Figure 11. Transport critical current $I_c$ properties of samples extracted from Spiral B, as compared to bare and insulated short samples. The $I_c$ distribution is homogeneous in all 6 layers of Spiral B, although lower than in the short bare and insulated samples, which have almost identical but higher $I_c$. In these samples reacted under 1 bar of pure $O_2$, we expect the longer length samples to have lower $I_c$ due to dedensification caused by internal gas pressure [17,18].

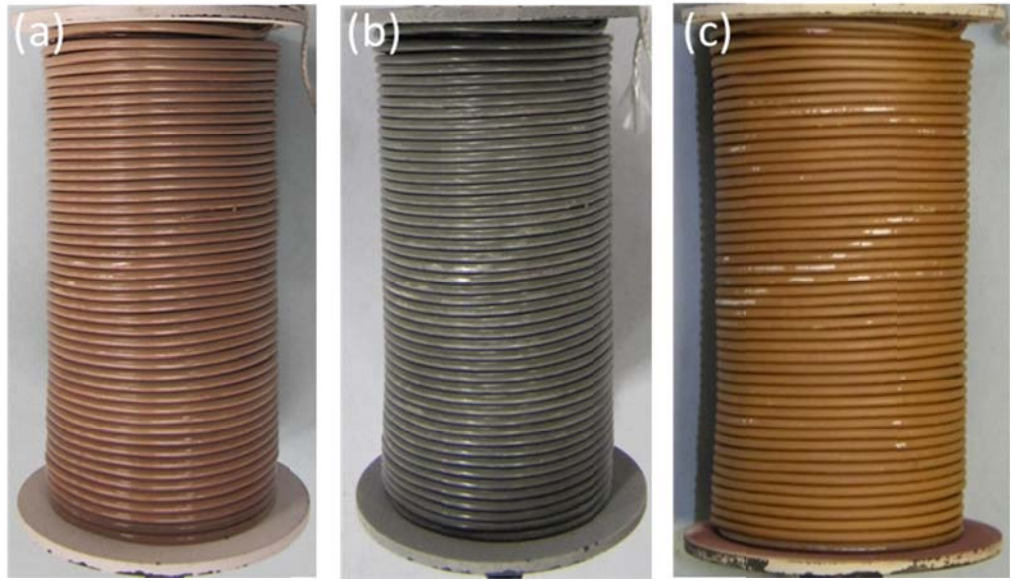

Figure 12. Insulated Bi-2212/Ag RW wound superconducting coil, seen (a) after winding, (b) after insulation burn-off and (c) after full 10 bar OP processing. Except for some insulation flaking resulting from improper handling during coil winding, the insulation was in good condition after full reaction and survived epoxy impregnation and full quench current testing in 31.2 T background field.

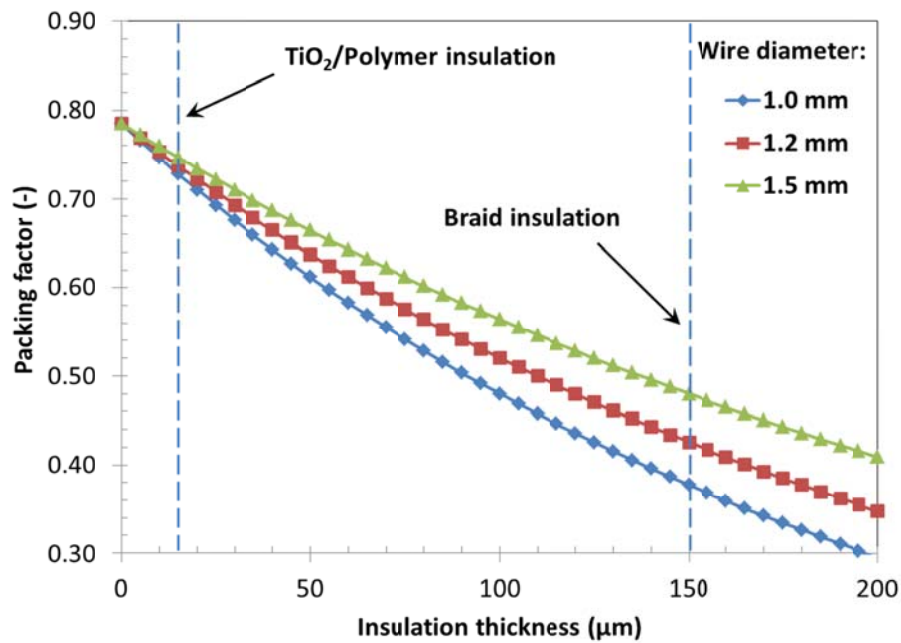

Figure 13. Comparison of packing factors for thin $TiO_2$/polymer insulation (15 μm) and alumino-silicate braid insulation (150 μm) for Bi-2212/Ag round wires with typical diameters of 1.0, 1.2 and 1.5 mm, respectively. Use of $TiO_2$/polymer insulation instead of alumino-silicate braid insulation allows a packing factor of ~0.74, as compared to 0.38-0.48 expected for the braid insulation for Bi-2212/Ag RW with a typical diameter of 1.0-1.5 mm.